\documentclass[twocolumn]{aastex7}

\newcommand\ugc{UGC 6578}
\newcommand\kms{km s$^{-1}$~}
\newcommand\sbunits{erg s$^{-1}$ cm$^{-2}$ arcsec$^{-2}$~}
\newcommand{\HI}{$\text{H}\,\text{\small I}$~}

\begin{document}

\title{The ISM-CGM Transition of a Star-forming Dwarf Galaxy Resolved in Ionized Gas Emission}

\author[orcid=0000-0002-6558-9894]{Chloe Neufeld} 
\affiliation{Department of Astronomy, Yale University, 219 Prospect Street, New Haven, CT 06511}
\affiliation{Center for Interdisciplinary Exploration and Research in Astrophysics (CIERA), Northwestern University, 1800 Sherman Avenue, Evanston, IL 60201, USA}
\email[show]{chloe.neufeld@yale.edu}

\author[orcid=0000-0002-8282-9888]{Pieter G. van Dokkum}
\affiliation{Department of Astronomy, Yale University, 219 Prospect Street, New Haven, CT 06511}
\affiliation{Dragonfly Focused Research Organization, LLC}
\email{pieter.vandokkum@yale.edu}

\author[orcid=0000-0002-7075-9931]{Imad Pasha}
\affiliation{Center for Interdisciplinary Exploration and Research in Astrophysics (CIERA), Northwestern University, 1800 Sherman Avenue, Evanston, IL 60201, USA}
\affiliation{Dragonfly Focused Research Organization, LLC}
\email{imad.pasha@dragonfly1000.com}

\begin{abstract}
The circumgalactic media (CGM) of dwarf galaxies are expected to be tightly coupled to their star formation, as low virial temperatures imply a CGM rich in cold accretion and shallow potential wells allow for more effective feedback. In this work, we use Keck/KCWI and KCRM to spatially resolve the ionized gas at the boundary between the interstellar medium (ISM) and CGM in dwarf galaxy \ugc\ ($M_* = 10^{7.8}\,M_\odot$), out to $\sim0.1\,R_{\rm vir}$. We discover a clear transition between the ISM and CGM at $0.033$$R_{\rm vir}$, where the H$\alpha$ surface brightness profile changes from an exponential to a shallow power law, marking the first such detection at $M_*/M_\odot < 10^8$. BPT diagnostics and ionization budget calculations indicate that the gas in the inner CGM is likely ionized by the galaxy's own star formation activity rather than shocks or the UV background. Spatially resolved emission maps also reveal a kinematically distinct, low-metallicity clump embedded in the inner CGM, and we discuss the possibility that this is a direct detection of ongoing gas accretion. This work serves as a pilot study for the Baryon Recycling in Dwarf Galaxies Emission Survey (BRIDGES), which will combine spatially resolved KCWI observations of the ISM-CGM interface in low-mass galaxies with wide-field observations of diffuse ionized gas out to the virial radius with the MOTHRA array.

\end{abstract}



\section{Introduction} 

The baryon cycle -- the flow of gas in and out of galaxies -- is thought to be a primary driver of long-term galaxy evolution. Gas accreted from the intergalactic medium (IGM) fuels star formation in the interstellar medium (ISM), while stellar and active galactic nuclei (AGN) feedback drive enriched material back out into the circumgalactic medium (CGM), where it may be recycled into subsequent iterations of star formation \citep[][]{Peroux:2020, Faucher-Giguere:2023, Wright:2024,Chen:2026}. The CGM may contain as much as half of a galaxy's baryons \citep{Tumlinson:2017}, and the long-term evolution of galaxies cannot be understood without characterizing the exchange of gas between a galaxy and this reservoir. 

Most of our observational understanding of the CGM comes from absorption-line and X-ray studies of $\sim$$L^*$ galaxies  \citep{Chen:2010,Werk:2014,Prochaska:2017,Anderson:2013,Bogdan:2013,Zhang:2024}. The CGM of low-mass galaxies has not been studied as extensively, and is expected to differ fundamentally. Low-mass galaxies have low virial temperatures that favor cold accretion, in which $\sim10^4$ K gas flows efficiently onto the galaxy rather than being shock-heated into a hot, pressure-supported halo \citep{Keres:2005,Lochhaas:2020,Crain:2023}. Cosmological simulations predict that a virialized inner CGM ($\sim0.1 R_{\rm vir}$) forms only above $M_{\rm halo} \sim 10^{12}M_\odot$, coinciding with the transition from bursty to steady star formation and the presence of rotation-dominated disks; below this halo mass, the inner CGM is expected to remain cool and dynamically disturbed \citep{Stern:2020}. The resulting supply of cold gas, together with systematically higher gas fractions than in more massive systems \citep{Peeples:2014}, may explain why nearly all isolated dwarf galaxies are actively star forming \citep{Geha:2012}. 

The shallow potential wells of low-mass systems also produce efficient stellar feedback, driving episodic outflows \citep{Mac_Low:1999,Muratov:2015,McQuinn:2019} and cyclical bursts of star formation rather than the smooth star formation histories of more massive systems \citep{Weisz:2012,Hopkins:2014, Sparre:2017,Emami:2019}. These outflows efficiently remove material, with low-mass galaxies retaining only a small fraction of the metals they have produced \citep{Peeples:2014,Chisholm:2018,Yu:2025}, implying that the CGM may hold a substantial fraction of their metal and baryon budget \citep{Zheng:2024,Piacitelli:2025}. Simulations predict that the result is a disturbed, clumpy, and extended CGM rather than the smooth virialized halo expected at higher masses \citep{Fielding:2017,Li:2021}. 

Observationally, however, both the extent of feedback-driven outflows in this mass regime and their effect on the CGM remain poorly constrained, as the CGM is faint in emission and has low column densities in absorption. Absorption-line studies have nonetheless shown that isolated dwarf galaxies host metal-enriched, multiphase gas in their halos \citep{Bordoloi:2014,Johnson:2017,Mishra:2024,Zheng:2024,Polzin:2026}, consistent with the efficient ejection of metals by outflows. Although detailed studies of individual systems are beginning to constrain the physical conditions and origins of this gas \citep{Fox:2026,Johnson:2026}, most
constraints come from averaging results from pencil-beam sight lines across large samples, which smooths over the clumpy, disturbed structure predicted for the dwarf CGM. 

Direct imaging of the gas distribution around dwarf galaxies
would resolve many of the open questions: it would enable us to see outflows, and potential accretion flows, for individual galaxies, and through kinematics and line ratios we could resolve the baryon cycle on a per-galaxy basis.
Deep 21\,cm observations have been highly informative, showing that dwarf galaxies often host \HI extending beyond their optical disks \citep{Begum:2008,Hunter:2012,Jaiswal:2020}.
However, this gas traces only the neutral component of a largely ionized medium and carries no information on its metallicity or ionization state, which are needed to distinguish enriched outflows from accreting gas \citep{Werk:2014,Fumagalli:2017}. Optical emission studies, which do detect this ionized gas, have largely focused on winds and diffuse ionized gas (DIG) within and immediately around the galaxy \citep{Hunter:1990,Hunter:1997,McQuinn:2019,Marasco:2023}. As a result, the spatial structure of the dwarf CGM, specifically in its innermost regions where gas is directly exchanging with the galaxy, remains largely unexplored.

Integral field spectroscopy with instruments such as Keck/KCWI and KCRM \citep{Morrissey:2018,McGurk:2024} has proven to be a viable method of mapping the resolved inner CGM of individual systems in optical emission \citep{Nielsen:2024,Lin:2025}, but these studies have thus far targeted galaxies with $M_*\gtrsim 10^9 M_\odot$. The Baryon Recycling In Dwarf Galaxies Emission Survey (BRIDGES) is designed to extend such measurements to lower masses, where simulations predict the largest departure from the CGM of massive galaxies. BRIDGES targets star-forming, low-mass ($\log \rm{M}_*/\rm{M}_\odot = 7-9.5$) galaxies to probe the relationship between feedback, star formation, and the CGM, combining (1) high spatial- and spectral- resolution observations of the host galaxies and their inner CGM using KCWI/KCRM on Keck II, and (2) deep, halo-scale narrowband emission maps of the dwarf CGM with the Modular Optical Hyperspectral Robotic Array (MOTHRA; \citealt{Lokhorst:2019, van_Dokkum:2026}), a facility currently under construction. Together, these data will connect the resolved star formation and outflow properties of each galaxy to the morphology, kinematics, metallicity, and ionization state of its CGM. Because bursty star formation is expected to imprint on the CGM with a time delay, such measurements may constrain the star formation duty cycle of dwarf galaxies in a way that their own emission cannot. The full sample will be described in a forthcoming paper (Neufeld et al., in prep).

In this pilot study, we establish the observational framework for BRIDGES by mapping a single nearby low-mass galaxy, \ugc\ ($M_*\approx 10^{7.8}\,M_\odot$), and its inner CGM in optical emission. \ugc\ is known to host an \HI reservoir extending beyond its star-forming disk as well as a low column density \HI cloud of uncertain origin \citep{Taylor:1995,van_Zee:1998,Jaiswal:2020}, making it a good initial target. Using KCWI/KCRM, we map its ionized gas out to $\sim0.1 R_{\rm vir}$ and address the presence of a distinct transition between the ISM and CGM, the ionization of the gas beyond this transition, the kinematics of the inner CGM in relation to the disk, and the origin of a bright H$\alpha$ clump in the galaxy's outskirts. 

This paper is organized as follows. In Section \ref{sec:methods} we describe our KCWI/KCRM observations and data reduction. We present our emission line maps, surface brightness profiles, kinematics, and metallicity diagnostics in Section \ref{sec:results}. In Section \ref{sec:discussion}, we discuss the implications for the location and nature of the ISM-CGM interface in low-mass galaxies, the ionization mechanism of the inner CGM, and the origin of the external clump. We summarize our conclusions and the future prospects of the BRIDGES survey in Section \ref{sec:conclusions}. Throughout, we assume a flat $\Lambda$CDM cosmology with H$_0$=70 km s$^{-1}$ Mpc$^{-1}$, $\Omega_{\rm{m}}$=0.3, and $\Omega_\Lambda$=0.7, and distances are quoted in physical units.

\section{Methods}\label{sec:methods}

\subsection{Target}
\ugc\ is a nearby, irregular, low-mass (M$_*$/ M$_\odot = 10^{7.8}$) galaxy in the BRIDGES sample (see full properties in Table \ref{tab:ugc6578}). The optical $r$-band image of the galaxy was obtained from the Legacy Survey \citep{Dey:2019} and was fit with the \texttt{pysersic} code \citep{Pasha:2023} using a single Sérsic profile. \ugc\ has a measured Sérsic index of 1.28$\pm0.01$, near exponential, with an effective radius of $7.98\pm0.03$ arcsec and $r_{80}$ of 15.49 arcsec, with $r_{80}$ computed using the fitting function provided in \citet{Miller:2019}. These radii correspond to 0.54 kpc and 1.05 kpc, respectively, at the distance of \ugc.

\ugc\ is already known to host an extended reservoir of neutral gas, making this a prime candidate for studies of its ionized gas distribution. Observations of \HI 21 cm-line emission from the Very Large Array \citep[VLA; ][]{Taylor:1995,van_Zee:1998} and Giant Meter-wave Radio Telescope \citep[GMRT; ][]{Jaiswal:2020} show a centrally peaked \HI distribution extending along the major axis and well beyond the star-forming regions of the galaxy, along with kinematically distinct clumps. High column density \HI coincides with regions of intense star formation, suggesting a recent starburst \citep{van_Zee:1998}. \citet{van_Zee:1998} also identify a low column density \HI cloud with an estimated mass of $\sim6\times10^6 M_\odot$, whose origin cannot be constrained from the \HI data alone; we return to this feature in Section \ref{sec:clump}.

\begin{deluxetable}{lcll}
\tablecaption{UGC 6578 Galaxy Properties.\label{tab:ugc6578}}
\tablehead{
\colhead{Property} & \colhead{Value} & \colhead{Units} & \colhead{Reference}
}
\startdata
R.A.  & 11h36m36.8372s &  &  \\
Dec   & $+$00d48m56.284s &  &  \\
$z_{\rm gal}$ & 0.003686 &  &  \\
Distance & 13.98 & Mpc & \citet{Tully:2016}, \citet{Ohlson:2024} \\
SFR & $0.08\pm0.01$ & M$_{\odot}$~yr$^{-1}$ & \citet{Jaiswal:2020} \\
\HI Mass & $2.35\pm0.12$ & $M_{{HI}}/10^8 M_{\odot}$ & \citet{Jaiswal:2020} \\
Stellar Mass & $7.83$ & $\log(M_{\ast}/M_{\odot})$ & \citet{Zhao:2013} \\
Halo Mass\tablenotemark{a} & $10.4$ & $\log(M_{\rm vir}/M_{\odot})$ &  \\
$R_{\rm vir}$\tablenotemark{b} & $75$ & kpc & \\
$n$ & 1.28$\pm0.01$ &  &  \\
$R_{\rm{eff}}$\tablenotemark{c} & 0.541 $\pm$ 0.002 & kpc & \\
$12 + \log({\rm O/H})$ & $8.08\pm0.03$ &  & \citet{Zhao:2013} \\
\enddata
\tablenotetext{a}{Estimated from the stellar mass using the $z = 0$ abundance matching relation of \citet{Behroozi:2013}.}
\tablenotetext{b}{Virial radius calculated from the halo mass using the \citet{Bryan:1998} relation.}
\tablenotetext{c}{Effective radius calculated using the Sersic profile obtained from \textsc{pysersic}.}
\end{deluxetable}

\subsection{KCWI/KCRM Observations}

\ugc\ was observed with Keck/KCWI and Keck/KCRM on 18 and 28 February 2025 (PID:Y157) in two configurations: one targeting the galaxy itself to map its star formation and ISM gas properties, and one to probe the inner CGM to $\sim0.1$ R$_{\rm{vir}}$. Both configurations used the large slicer with the default $2\times2$ on-chip binning, giving a field of view of $33 ''\times20.4 ''$ and a spaxel size of $0.29 ''\times1.35 ''$. 

For the galaxy pointings, we used the BH3 grating centered at 5000 \AA{} (R$\sim$4500), and the RH2 grating centered at 6800 \AA{} (R$>$3250). Multiple pointings were observed to fully map out the central region of the galaxy, with exposures ranging from 1s to 600s (100s to 300s for KCRM) to balance saturating emission in the galaxy center and faint emission out to 2 kpc from the galaxy. Separate sky exposures were obtained for pointings that did not have sufficient in-frame sky coverage. 

For the inner CGM pointings, we used the BL grating centered at 4550 \AA{} (R$\sim$900) and RM1 grating centered at 7000 \AA{} (R$\gtrsim$1400). Pointing placement was guided by \HI emission data, with two pointings within R$_{\rm{HI}}=5.13$ kpc \citep[the radius at which \HI mass surface density falls to 1 M$_\odot$ pc$^{-2}$ as defined in][]{Jaiswal:2020} and one falling partially outside this radius. We repeated exposures of 990s (300s for KCRM) five to six times per pointing, and five sky frames outside the virial radius were taken throughout the night. 

\subsection{Data reduction}
We first remove cosmic rays in the raw data using the CRCombine package\footnote{\url{https://github.com/prappleizer/CRCombine}}. For each pointing, the code median-combines the individual exposures, subtracts this median image from each frame, and identifies cosmic rays as outliers in the residuals. The cosmic-ray masked frames are then mean-combined. The data were then reduced using the Python version of the KCWI Data Reduction Pipeline (DRP)\footnote{\url{https://github.com/Keck-DataReductionPipelines/KCWI_DRP}} with standard settings, with sky subtraction turned off. The central and CGM pointings each require additional post-processing steps, as detailed below, before cubes are aligned and coadded. 

For the central pointings, emission lines saturate in the longest exposures, and we thus combined exposures of different lengths to take this into account. We first search for saturated pixels in the frame with the longest exposure time and replace these with pixels from the next corresponding longest exposure that is not saturated, scaling the flux to account for the difference in exposure time. 

The faint CGM emission requires two steps for careful treatment of the sky background. First, we manually correct for a known, wavelength-dependent illumination gradient across the KCWI FOV that remains after DRP processing \citep[e.g.,][]{Rupke:2019}, following the method of \citet{Nielsen:2024}. In each of the five sky exposures, we mask sources by $3\sigma$-clipping the white light flux along each slice, then compute the mean spectrum of each slice and of the full cube. We smooth both with a 201-pixel boxcar and resample them onto a 40\,\AA{} grid. The ratio of each slice's spectrum to the cube's mean spectrum, averaged over the five sky exposures, gives a smooth wavelength-dependent response for each slice, which we divide out of all CGM science and sky cubes. We assume that this illumination gradient is stable across the night for the single configuration used, and we find no gradient in the KCRM data, so we skip this step for the red channel. 

After correcting for the illumination gradient, we collapse each sky exposure to a mean spectrum over its source-free spaxels, masked as above, and model the mean spectrum of each science frame as a linear combination of these five sky spectra. To allow the sky to vary with wavelength between exposures, each sky frame's coefficient is a piecewise-linear function of wavelength, with bins spaced every 35 (75) unmasked pixels in the KCWI (KCRM) cubes ($\sim$30\,\AA{}). We fit all coefficients simultaneously by linear least squares, so the sky model is continuous across the bins. We exclude $\pm7$\,\AA{} around each emission line at the galaxy's redshift from the fit and subtract the best-fit sky spectrum from every spaxel of the science cube. 

We use the Python package CWItools \citep{OSullivan:2020} to trim data cubes, perform WCS corrections, and align and coadd the data cubes for each pointing. We perform astrometric alignment using point sources present in each frame and use a drizzle value of 1.0, taking advantage of slight dithers between exposures, to obtain square $0.29 ''\times0.29 ''$ spaxels in the final coadded data cubes. 

\subsection{Emission line measurements}\label{sec:lines} 
We first correct each data cube for Milky Way foreground extinction using the \citet{Cardelli:1989} extinction law with $R_V=3.1$ and $E(B-V)=A_V/R_V=0.022$, with $A_V=0.068$ mag taken from \citet{Schlafly:2011} via NED. 

The CGM is inherently faint and diffuse; we thus optimally bin the data to maximize signal-to-noise ratios in the emission lines  H$\beta$, [OIII], H$\alpha$, and [NII]. We measure the signal-to-noise ratio in each unbinned spaxel first by subtracting the continuum level from the spectrum. For the blue side, we fit the continuum  in an 80\,\AA{} wide region around the line, and for the red side we measure the median continuum flux level in a region free of emission lines at $\lambda=6660$\,\AA{}. We subtract the measured continuum level and estimate the noise from the standard deviation of the same wavelength range. For each emission line, we sum the total flux in a 6\,\AA{} wide bin around the galaxy's redshift to capture any variation in the velocity structure of the ionized gas in these regions, without fitting a Gaussian to the inherently low signal-to-noise data (see Section \ref{sec:kinematics} for a more in depth discussion of velocity structure). We then Voronoi bin the cubes with VorBin \citep{Cappellari:2003,Cappellari:2012}, targeting a minimum signal-to-noise ratio of seven for each emission line. 

After binning, we remeasure the flux and its uncertainty from each bin's summed, continuum-subtracted spectrum. We estimate the uncertainty by integrating the flux in randomly placed 6\,\AA{} windows drawn from line-free regions of the same spectrum, bootstrapping over 2000 such windows. For [NII] in the CGM pointings, where the line is undetected, we take the 95th percentile of the same window distribution as the upper limit on the flux.

For the central pointings where H$\beta$ is detected at sufficiently high signal-to-noise ratio, we correct for internal extinction using the Balmer decrement, assuming an intrinsic H$\alpha$/H$\beta$ ratio of 2.86 \citep[Case B recombination;][]{Osterbrock:2006} and adopting the \citet{Cardelli:1989} curve with $R_V=3.1$. Because H$\alpha$ and H$\beta$ are Voronoi-binned independently, we derive the decrement as a function of galactocentric radius rather than bin by bin, using the ratio from the surface brightness profiles for both lines. In the outskirts of \ugc, the H$\beta$ signal is insufficient for resolved Balmer decrement measurement, so we instead apply a constant extinction correction. We assume $A_V$=0.03 for the CGM, from the average V-band extinction of the CGM in the local Universe, as estimated in \citet{Menard:2010}. 

Finally, we convert the extinction-corrected fluxes to H$\alpha$ surface brightnesses by dividing by the area of each bin. The 3$\sigma$ H$\alpha$ surface brightness limits per Voronoi bin are $8.5\times10^{-17}$\,\sbunits for the central galaxy pointings
and $2.5\times10^{-18}$\,\sbunits for the CGM pointings. Radial profile points average many bins and reach correspondingly fainter limits.

\section{Results: Mapping the Inner CGM}\label{sec:results}

\subsection{Spatial Distribution of Ionized Gas}
\begin{figure*}
    \centering
    \includegraphics[width=0.7\linewidth]{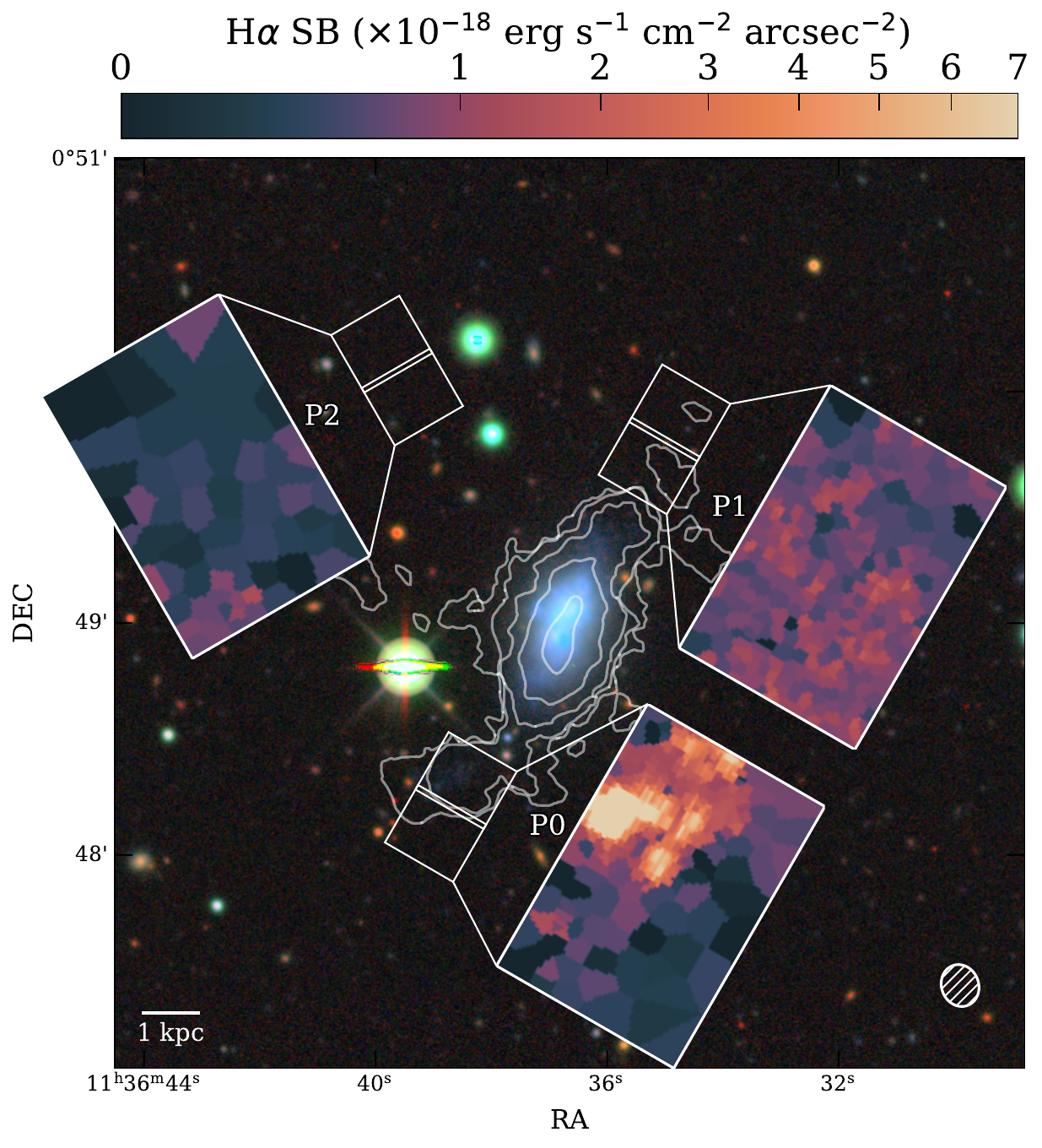}
    \caption{KCWI pointings for observations of UGC 6578 overlaid on optical imaging from the Legacy Surveys / D. Lang (Perimeter Institute), with H$\alpha$ surface brightness maps and the corresponding colorbar shown for the CGM pointings. Contours show the VLA natural-weight \HI column density of \ugc\ from \citet{van_Zee:1998}, digitized from their Figure~8b; contour levels are 2, 4, 8, 16, and $32\times10^{20}$\,cm$^{-2}$. The \HI beam ($11\farcs0\times9\farcs6$) is shown at the bottom right. We additionally show a scale bar indicating 1 kpc at the distance of \ugc.}
    \label{fig:halphamaps}
\end{figure*}

We present KCRM H$\alpha$ surface brightness maps overlaid on optical imaging of \ugc\ in Figure \ref{fig:halphamaps} and VLA \HI emission data shown as contours \citep{van_Zee:1998}. Faint H$\alpha$ is detected in all three inner CGM pointings, out to 6 kpc ($\sim$10\,\% of the virial radius). The morphology of the ionized gas is mostly smooth and volume filling, with the exception of the southern pointing (P0), around 3 kpc from the galaxy, where we see clumpy substructure with brighter H$\alpha$ detections (surface brightness peaking at $2\times10^{-17}$\sbunits). This external clump is also detected in GALEX FUV observations, with a flux of F$_\lambda$= 1.88$\times 10^{-16}$ erg s$^{-1}$ cm$^{-2}$ \AA$^{-1}$, and is well correlated with the \HI features discussed in \citet{van_Zee:1998} and \citet{Jaiswal:2020}. We discuss the possible origins of this external clump in Section \ref{sec:discussion}.

\subsection{H$\alpha$ surface brightness profile}

\begin{figure*}
    \centering
    \includegraphics[width=\linewidth]{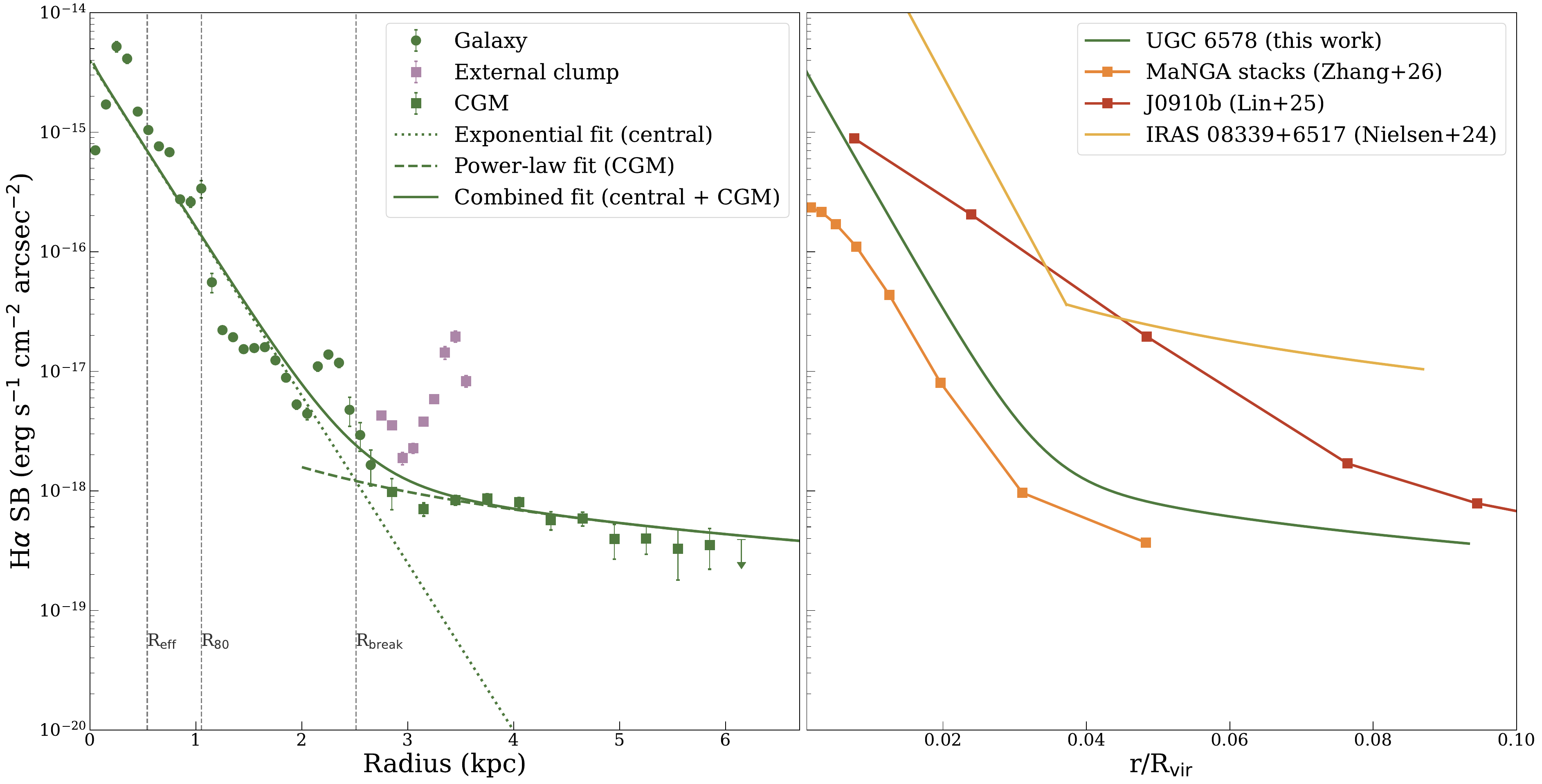}
    \caption{H$\alpha$ surface brightness profiles. Left: H$\alpha$ vs. radius from the center of the galaxy in kpc, where points are the mean surface brightness values in bins of 0.1 kpc for the central (green circles) and external clump (purple squares) data and 0.3 kpc for the CGM pointings (green squares). Points with surface brightness values below 2$\sigma$ are plotted as upper limits. The dotted green line shows the exponential fit to the galaxy spaxels, and the dashed green line shows the power law fit to the CGM spaxels (excluding the external clump). Vertical lines for $R_{\rm eff}$, $R_{80}$, and the radius at which the fits to the galaxy and CGM component cross ($R_{\rm break}$) are also shown. Right: The two component surface brightness profile for UGC 6578 from this work (green), along with surface brightness profiles from three other studies, for more massive galaxies. These profiles are scaled by the virial radius of the respective galaxies.}
    \label{fig:sbprofile}
\end{figure*}

We show the H$\alpha$ surface brightness profile as a function of radius from the center of the galaxy in Figure \ref{fig:sbprofile}. The radii containing 50\,\% and
80\,\% of the optical light are indicated in Figure\ \ref{fig:sbprofile} with vertical dashed lines. We additionally plot $R_{\rm break}$: the radius at which we find a pronounced break in the surface brightness profile. Following \citet{Nielsen:2024}, we fit the emission from the galaxy itself (green circles in the left panel) to an exponential profile, shown as the dotted curve, with the form $ae^{-r/r_{0,\rm exp}}$, where $a=3.9\times10^{-15}$ \sbunits is the central surface brightness and $r_{0,\rm exp}=0.31$ kpc is the scale radius. Beyond the break radius the profile flattens, and we associate this outer gas with the CGM of \ugc. We indicate this flat part of the profile with green squares, and fit it with a powerlaw (dashed green line) with the form $10^{-18}(r / r_{0,\rm pow})^k$, where $r_{0,\rm pow}=2.9$ kpc is the radius at which the fit reaches a surface brightness of $10^{-18}$\sbunits and $k=-1.18$ is the slope. The transition between the exponential profile (galaxy) and power law profile (CGM) occurs at $R_{\rm break}\sim 2.5$\,kpc, or roughly 0.033\,$R_{\rm vir}$. The bright H$\alpha$ clumps at $\sim$3 kpc in the southern pointing (P0) are computed and plotted separately (purple squares) as they have elevated flux over the smooth component. 

Additional surface brightness profiles from recent studies are plotted as lines on the right panel in Figure \ref{fig:sbprofile} in terms of virial radius, along with the combined exponential and power-law fit to our data. These relations are as follows. \citet{Zhang:2026} provide median stacked H$\alpha$ profiles of $z\sim0.015-0.146$ MaNGA galaxies, with  a large range of masses; we plot their data for galaxies with $-1.75 <\log~{\rm SFR}/M_\odot\,\rm{yr}^{-1}<-1$ to correspond to the SFR of \ugc, and as their surface brightness is in terms of effective radius, we scale their provided radii by the effective and virial radii of \ugc\ (0.54 kpc and 75 kpc, respectively). \citet{Nielsen:2024} provide the H$\beta$ surface brightness profile of a $10^{10}M_\odot$ stellar mass galaxy, which we scale by a factor of 2.86 to correspond to our H$\alpha$ profile, and we divide by the virial radius reported for their target galaxy (115 kpc). \citet{Lin:2025} provide data for the H$\alpha$ surface brightness profile of a $10^{9.2}M_\odot$ galaxy, which we again rescale according to the virial radius reported for their target galaxy (103 kpc). We discuss this comparison in Section \ref{sec:transition}.

\subsection{Ionization mechanism and metallicity of the inner CGM}\label{sec:ionization}
\begin{figure*}
    \centering
    \includegraphics[width=\linewidth]{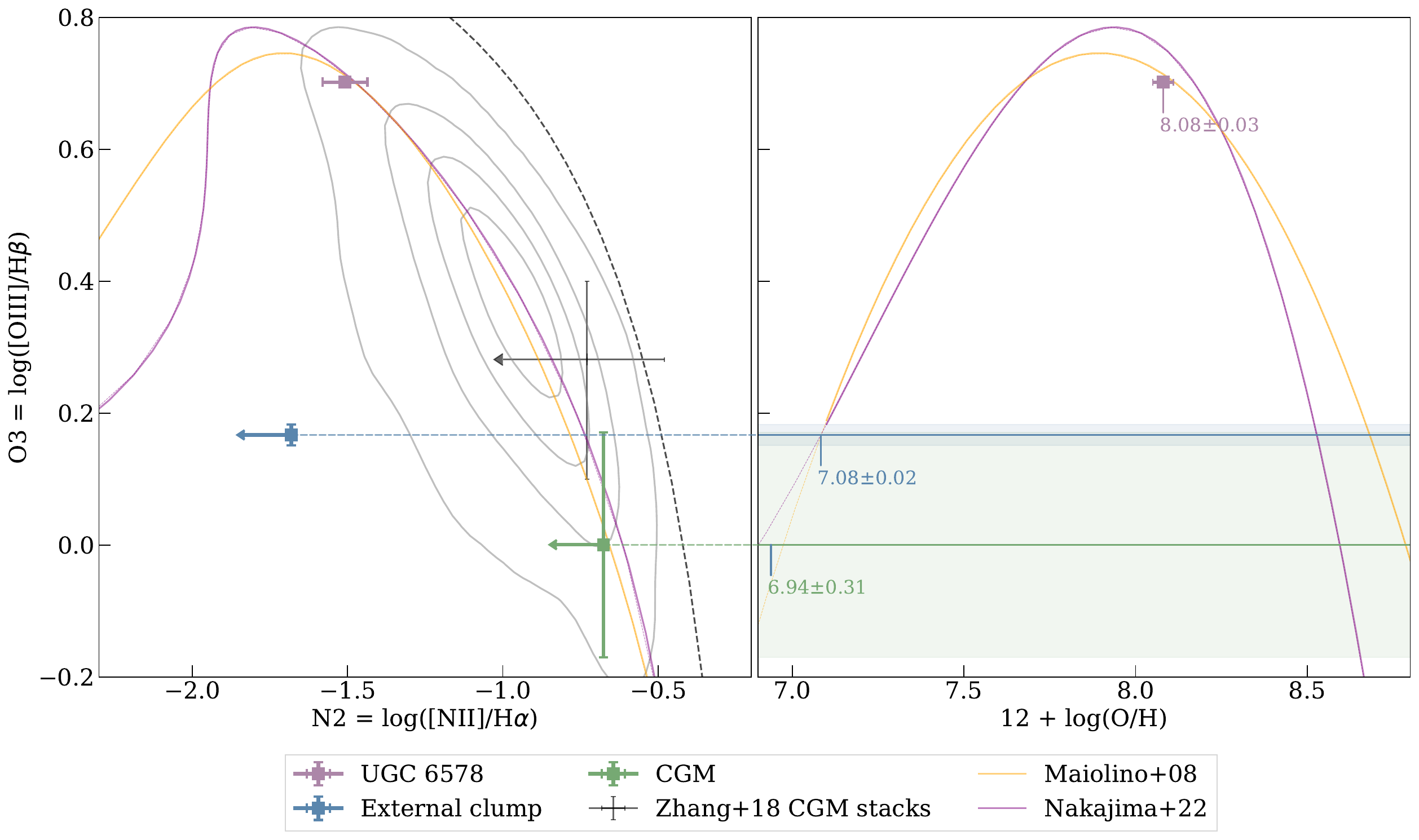}
     \caption{Left: BPT diagram for the galaxy (purple square), external clump (blue square), and stacked CGM pointings (green square). We include two relations from the literature, which are used to estimate metallicity from strong-line diagnostics like O3. We also plot a sample of SDSS low-mass galaxies for reference as the contour lines, as well as the result from \citet{Zhang:2018} for stacked measurements of the CGM of low-mass galaxies. The dashed line is the delineator between HII and AGN regions on typical BPT diagrams from \citet{Kauffmann:2003}. Horizontal dashed lines connect each of the \ugc\ measurements with the panel on the right: O3 vs. metallicity. We include the same relations as in the left panel, with dashed lines indicating where the relations were extrapolated to lower metallicities. We plot the galaxy at our measured O3 value and the metallicity found by \citet{Zhao:2013}, and we plot our measured O3 values for the external clump and CGM as horizontal lines. }
    \label{fig:bpt}
\end{figure*}

To probe the ionization source of the gas in the outskirts, we calculate the line ratios N2 = log([NII]/H$\alpha$) and O3 = log([OIII]/H$\beta$) for the galaxy, the external clump, and the combined CGM pointings (excluding the clump), and plot them on the Baldwin–Phillips–Terlevich (BPT) diagram \citep{Baldwin:1981} in the left panel of Figure \ref{fig:bpt}. The dashed distinguishing line comes from \citet{Kauffmann:2003}, separating ionization mechanisms from shocks or AGN and HII regions \citep[e.g.,][]{Kewley:2019}. For comparison, we show low-mass galaxies (M$_*=10^{7-9}M_\odot$) from the NSA catalog \citep{Blanton:2011} (grey contours) and stacked CGM measurements of low-mass galaxies (M$_* < 10^{10.4}M_\odot$) at 10 to 50 kpc radii from \citet{Zhang:2018}. Errorbars are derived from 1$\sigma$ noise values in the spectra. 

We detect H$\alpha$ and [OIII] in all three CGM pointings, but [NII] is undetected outside of the galaxy and H$\beta$ is detected only in the clump. For the CGM, we therefore adopt the [NII] upper limit described in Section \ref{sec:lines}, and we estimate H$\beta$ using H$\alpha$ and the standard line ratio of H$\beta$/H$\alpha$ = 0.3, following \citet{Zhang:2018}. As extinction in the CGM is expected to be low \citep{Zaritsky:1994, Menard:2010}, this provides a lower limit on H$\beta$ and thus an upper limit on O3. With both ratios as upper limits, this places the CGM firmly in the star-forming region of the BPT diagram. The galaxy, clump, and CGM are all consistent with stellar photoionization, as are the \citet{Zhang:2018} stacks. 

At low metallicity, these ratios also depend on metallicity, so we additionally compare our O3 measurements to strong-line metallicity calibrations on the right panel of Figure \ref{fig:bpt} \citep{Maiolino:2008,Nakajima:2022}. The galaxy's O3 of 0.7 and its direct-method metallicity of 12+log(O/H) = 8.08 $\pm$ 0.03 \citep{Zhao:2013} are consistent with both relations near their peak. The O3 calibrations are double-valued, but the clump's N2 upper limit (-1.68) lies below the galaxy's value, placing it on the lower-metallicity branch. Its O3 of $0.17\pm0.02$ then implies 12+log(O/H) $\approx 7.1$, about 1 dex below the galaxy. For the combined CGM pointings, the N2 limit does not break this degeneracy, but assuming the lower branch, O3 implies 12+log(O/H) $\approx 6.9\pm0.3$. 

\subsection{Kinematics}
\label{sec:kinematics}
We measure H$\alpha$ velocities from the peak values of the H$\alpha$ line profiles. The kinematic structure of the ionized gas in the galaxy 
is consistent with that found in the neutral gas distribution \citep{van_Zee:1998,Jaiswal:2020}, with a velocity gradient across the semi-major axis of the galaxy spanning from 1040 km/s to 1120 km/s. To improve the SNR in the smooth CGM pointings, we collapse each pointing to a single bulk velocity. P1, on the major axis, has a bulk velocity of 1066 \kms, consistent with disk rotation on the approaching side of the galaxy. This suggests that the inner CGM along the major axis co-rotates with the disk. P2, aligned with the minor axis, has a bulk velocity of 1080 \kms, nearer to the systemic velocity of the galaxy, implying it is not some form of wind or outflow. 

The external clump in P0 is kinematically distinct from both the disk and the smooth CGM (Figure \ref{fig:clump}), which \HI data were unable to resolve. Its velocities span $\sim1060-1140$ \kms, a range as wide as the rotation of the galaxy itself, with a coherent gradient across its brightest component.

\section{Discussion}\label{sec:discussion}
\subsection{The ISM-CGM transition}\label{sec:transition}
The innermost region of the CGM marks the interface between the ISM and the CGM, where outflows are launched and gas is accreted onto the galaxy \citep{Norman:1989,Putman:2012}. Because dwarf galaxies host extended \HI disks, the location of this interface is not inherently obvious; it may not coincide with the edge of the stellar disk or with a fixed \HI column density contour. We therefore define it empirically as the radius at which the ionized gas stops following the galaxy's own emission profile and instead behaves as a distinct component, with its own ionization and kinematic properties. In more massive galaxies, such a transition has recently been identified \citep{Nielsen:2024,Lehner:2026,Zhang:2026}. In \citet{Nielsen:2024}, it appears as a change in the slope of the ionized gas surface brightness profile accompanied by a change in the dominant ionization mechanism. At these masses, accreting gas is thought to join the disk at its outskirts, often after hot halo gas cools under angular-momentum support \citep{Stewart:2011,Fernandez:2012,Trapp:2022,Hafen:2022}. Dwarf galaxies lack this hot inner halo \citep{Stern:2020}, so whether a distinct boundary exists in low-mass systems, where it is located, and what ionizes the gas beyond it are all open questions.

The flattening of the H$\alpha$ surface brightness profile at $R_{\rm break}\approx2.5$ kpc ($0.033 R_{\rm vir}$) strongly suggests that, in the dwarf galaxy \ugc, the inner CGM is a distinct component, not simply an extension of the galactic disk. This is the first clear detection of the ISM-CGM boundary in a galaxy with stellar mass $<10^8M_\odot$, and there is tentative agreement with simulations of the CGM of dwarf galaxies (Neufeld \& Piacitelli, in prep), as well as simulations of Milky Way-mass halos \citep{Saeedzadeh:2026}. 

Our results are broadly consistent with recent CGM observations of this ISM-CGM transition in more massive galaxies (Figure \ref{fig:sbprofile}, right). The profile of \citet{Nielsen:2024} and the MaNGA stacks of \citet{Zhang:2026} both show a distinct break in slope. Despite two orders of magnitude difference in stellar mass, the transition radius of \ugc\ is comparable to that found by \citet{Nielsen:2024} of $0.04R_{\rm vir}$. This suggests that there may be a universal transition at the disk edge of galaxies, independent of stellar mass, but a larger sample is needed to confirm this trend and to constrain how the transition radius varies with stellar mass.

The kinematics of the inner CGM also support this picture. Along the major axis, the ionized gas beyond $R_{\rm break}$ co-rotates with the disk (Section \ref{sec:kinematics}). Co-rotating cool gas is commonly seen along the major axes of more massive galaxies \citep{Kacprzak:2010,Ho:2017,Martin:2019,Zabl:2019}. Our observations suggest that, even without a hot inner halo, gas at the ISM-CGM interface of a dwarf galaxy retains angular momentum aligned with the disk. 

\subsection{Ionization of the inner CGM}

The observed H$\alpha$ surface brightness in the inner CGM is significantly brighter than expected from the extragalactic UV background (UVB) alone. The UVB at $z=0$ predicts $1.2\pm0.5\times10^{-19}$\sbunits \citep{Fumagalli:2017}, while we measure a mean surface brightness in the inner CGM of around $6\times10^{-19}$\sbunits, which is a factor of $\sim5$ brighter. \citet{Fumagalli:2017} also find that ionization from star formation in a galaxy drops where N$_{\rm{HI}}<10^{19}$ cm$^{-2}$; our pointings are within this boundary, implying that stellar ionization is expected in the inner CGM region that we are probing here. Results from BPT diagnostics in Section \ref{sec:ionization} serve as further evidence that the ionization mechanism of the gas in the inner CGM is indeed stellar, in agreement with emission-line stacks of \citet{Zhang:2026}, who find that the SFR of normal galaxies is sufficient to ionize their inner CGM. 

This contrasts with \citet{Nielsen:2024}, who find that the CGM around the ISM-CGM transition radius of a more massive galaxy is shocked, and that beyond 6 kpc ($\sim$0.05$R_{\rm vir}$), the gas undergoes another transition where it becomes ionized by a non-shock mechanism such as the UVB. 
This difference may reflect the lower virial temperature of \ugc\ or difference in its ionization budget. We also probe only the inner $\sim 0.1 R_{\rm vir}$ and may not reach far enough into the CGM to observe a shift to UVB- or accretion-dominated ionization. 

To test whether the galaxy's star formation can supply the required ionizing photons, we estimate the ionizing photon rate needed to produce the observed level of H$\alpha$ emission following \citet{Lokhorst:2022} --- we use the Case B relation from \citet{Osterbrock:2006} and assume ionization equilibrium, giving the number of ionizing photons for optical line recombination emission as

\begin{equation}
    Q_{\rm{H}^0} = \frac{L_{\rm{H}\alpha}}{h\nu_{\rm{H}\alpha}} \frac{\alpha_{\rm{B}}}{\alpha_{\rm{H}\alpha}^{\rm{eff}}}
\end{equation}

where $L_{\rm{H}\alpha}$ is the H$\alpha$ luminosity, $\alpha_{\rm{B}}= 2.59 \times10^{-13}\rm{cm}^3\rm{s}^{-1}$ is the recombination coefficient for Case B at T=10$^4$K, and $\alpha_{\rm{H}\alpha}^{\rm{eff}}=11.7\times10^{-14}\rm{cm}^3 \rm{s}^{-1}$ is the H$\alpha$ effective recombination coefficient at T=10$^4$K and electron density $n_e\approx 1\rm{cm}^{-3}$ \citep{Osterbrock:2006}. We only consider the CGM pointings P1 and P2 in this case, as the clump in P0 could be actively forming stars and thus may contribute to ionization of the gas around it. The combined luminosity of these two pointings is $4.5\times10^{36}$ erg s$^{-1}$, and the total ionizing photon rate needed to sustain the observed H$\alpha$ emission is thus $3.3\times10^{48}$ photons s$^{-1}$. We then estimate the total ionizing photon rate from the galaxy itself, using the \citet{Kennicutt:1998a}  relation:

\begin{equation}
    L_{\rm{H}\alpha} [\rm{erg/s}] = (1.26\times10^{41}) \times \rm{SFR} [\rm{M}_\odot\rm{/yr}]
\end{equation}

where SFR = 0.08 M$_\odot$ yr$^{-1}$ \citep[][also recalculated and confirmed using our KCRM H$\alpha$ data and the \citet{Kennicutt:1998a} relation]{Jaiswal:2020}. The ionizing flux from the galaxy itself is thus $7.4\times10^{51}$ photons s$^{-1}$. 

The galaxy produces $\sim2000$ times more ionizing photons than are needed to sustain the H$\alpha$ emission in P1 and P2. Our pointings capture only a small fraction of the inner CGM, but even if its total H$\alpha$ luminosity is $\sim$100 times that in our pointings, an escape fraction of only a few percent would suffice. This is comparable to the $3.3\pm0.7$\,\% measured for the local dwarf Haro 11 \citep{Leitet:2011}, so the galaxy's own star formation can plausibly power the observed emission. 

The inner CGM of this dwarf galaxy is thus structurally similar to more massive systems, with a possibly universal ISM-CGM transition radius. However, the ionization mechanism beyond this transition is distinct, as it is dominated by the galaxy's own stellar ionization rather than shocks or the UVB, consistent with a lower virial temperature and a radiation field that remains significant at these radii.

\subsection{Origin of the external clump}\label{sec:clump}
\begin{figure*}
    \centering
    \includegraphics[width=\linewidth]{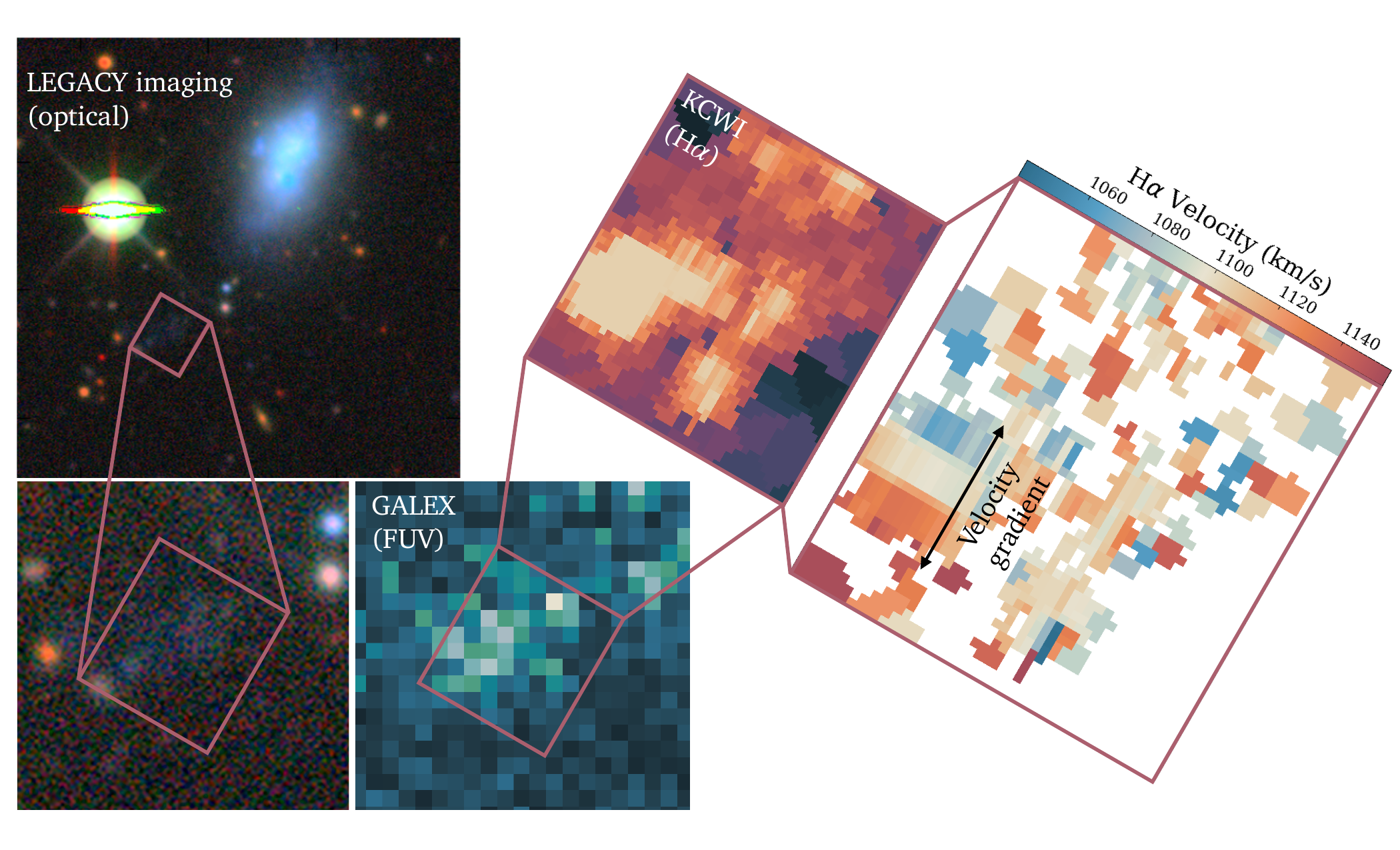}
    \caption{The external clump as viewed from different data products: optical imaging from the Legacy Surveys and a zoom-in on the clump, GALEX FUV imaging, KCWI H$\alpha$ surface brightness maps from this work, and velocity maps derived from the H$\alpha$ emission with a corresponding colorbar. We denote the location at which we observe a velocity gradient in this zoom-in region with an arrow located along the gradient, corresponding to the bright H$\alpha$ clump in the preceding panel.}
    \label{fig:clump}
\end{figure*}

The bright H$\alpha$ clump in P0 is physically distinct from both the galaxy disk and the smooth CGM component. It is clumpy, has a coherent velocity gradient spanning a range as wide as the disk rotation itself, and has a metallicity $\sim$1 dex below that of \ugc. Figure \ref{fig:clump} shows the clump in different tracers. In optical imaging from the Legacy Surveys \citep{Dey:2019}, the clump is faint and distinct from the main disk. It is, however, detected in GALEX FUV imaging \citep{Martin:2005}, indicating active star formation within the clump itself, so its gas is not ionized solely by the galaxy's star formation. A zoom in on the H$\alpha$ surface brightness map shows the morphology of the clump in ionized gas, and kinematics from the H$\alpha$ emission indicate a distinct rotation curve associated with the brightest component. 

The clump coincides with the low column density  cloud identified by \citet{van_Zee:1998}, who posit that this irregular gas could be outflow ejecta from previous star formation, tidal debris, or material being accreted onto the galaxy. With an estimated mass of $\sim6\times10^6$ M$_{\odot}$, only $\sim$3\,\% of the total \HI in the system, the cloud is unlikely to be a companion galaxy \citep{van_Zee:1998}, and although the \HI shows a tail-bridge morphology, no interaction partner has been identified near \ugc\ \citep{Taylor:1995,Jaiswal:2020}. The resolution of the \HI data was insufficient to reveal the clump's internal kinematics. 

Our data can help constrain these scenarios. An outflow from a starburst in the main galaxy is disfavored by the low metallicity, as enriched ISM gas blown out of the galaxy should be at least as metal-rich as the galaxy itself. An infalling, extremely low-mass dwarf galaxy would naturally explain the coherent rotation, low metallicity, and in situ star formation, but requires an independently bound object, for which the \HI data provide no evidence. An infalling pristine gas cloud is consistent with the low metallicity, and shearing as it reaches the disk could produce the clumpy morphology. Its coherent velocity gradient is harder to explain, but the cloud may retain angular momentum from its infall trajectory. Compression at the disk-CGM interface could then trigger the star formation traced by the FUV emission. 

Ambiguities on the formation scenario remain: line-of-sight kinematics do not provide the full 3D velocity of the gas, limiting what we can say about the rotation or orbit of the clump. We also cannot definitively rule out a gas-rich, optically faint satellite without deeper and higher resolution \HI data targeting the clump specifically. 

Either way, this distinct clumpy component illustrates that the ISM-CGM boundary is an active site of baryon cycling: infalling gas reaching this interface can trigger star formation, contributing to disk growth from the outside in. The fact that this object appears at the disk edge that we determined from the H$\alpha$ surface brightness profile break indicates that this is an extremely important regime for future study.

\section{Conclusions}\label{sec:conclusions}

In this work, we used Keck/KCWI and KCRM to map the ionized gas at the ISM-CGM interface of the dwarf galaxy \ugc\ ($M_*=10^{7.8}\,M_\odot$) in optical emission, out to $\sim0.1\,R_{\rm vir}$. These observations show that the ISM-CGM interface is a dynamic region with distinct structures visible in ionized gas emission. Our main results are as follows:

\begin{enumerate}
    \item A distinct ISM-CGM transition: The H$\alpha$ surface brightness profile changes from an exponential to a shallow power law at $R_{\rm break}\approx 2.5$\,kpc ($0.033\,R_{\rm vir}$). This is the first detection of the ISM-CGM transition in a galaxy with $M_* \lesssim 10^8\,M_\odot$. This transition radius is comparable to that found in more massive galaxies, indicating that the ISM-CGM interface may occur at a similar fraction of the virial radius across stellar mass, though a larger sample is needed to test this. 
    \item The inner CGM is photoionized by the galaxy's own star formation: The inner CGM is $\sim$5 times brighter in H$\alpha$ than expected from the UV background alone, and its line ratios place it in the star-forming region of the BPT diagram. The star formation in \ugc\ produces $\sim$2000 times the ionizing photons needed to sustain the observed emission, so an escape fraction of only a few percent would suffice. This contrasts with the shock-ionized transition region found in a more massive galaxy \citep{Nielsen:2024}, consistent with the lower virial temperature of a low-mass halo. 
    \item A spatially, kinematically, and chemically distinct clump in the inner CGM: The clump has a coherent velocity gradient, a metallicity $\sim$1\,dex below that of \ugc, and in situ star formation traced by FUV emission. Its low metallicity disfavors an origin in galactic outflows. We argue that this is either an infalling gas clump being sheared onto the disk of the galaxy or an infalling, extremely low-mass dwarf galaxy; either scenario would make it a direct detection of ongoing accretion at the ISM-CGM interface.
\end{enumerate}

This pilot study provides a first look at the innermost CGM of a dwarf galaxy, including the transition region between the ISM and CGM. However, the small field of view of current IFU instruments limits us to 10\,\% of the virial radius. Studying the accretion and ionization mechanisms of the outer CGM requires a larger field of view and deeper observations of the ionized gas. The wide field of view of the upcoming MOTHRA facility will allow us to probe the entire CGM of \ugc, out to the virial radius and beyond. The full baryon cycle and connections to the inner galaxy across a sample of dwarf galaxies at different stages of their evolution, the ultimate goal of the BRIDGES program, will be discussed in forthcoming papers. This paper serves as the first step towards a full baryon census in low-mass galaxies.

\begin{acknowledgments}
Some of the data presented herein were obtained at Keck Observatory, which is a private 501(c)3 non-profit organization operated as a scientific partnership among the California Institute of Technology, the University of California, and the National Aeronautics and Space Administration. The Observatory was made possible by the generous financial support of the W. M. Keck Foundation. The authors wish to recognize and acknowledge the very significant cultural role and reverence that the summit of Maunakea has always had within the Native Hawaiian community. We are most fortunate to have the opportunity to conduct observations from this mountain.

The Legacy Surveys consist of three individual and complementary projects: the Dark Energy Camera Legacy Survey (DECaLS; Proposal ID \#2014B-0404; PIs: David Schlegel and Arjun Dey), the Beijing-Arizona Sky Survey (BASS; NOAO Prop. ID \#2015A-0801; PIs: Zhou Xu and Xiaohui Fan), and the Mayall z-band Legacy Survey (MzLS; Prop. ID \#2016A-0453; PI: Arjun Dey). DECaLS, BASS and MzLS together include data obtained, respectively, at the Blanco telescope, Cerro Tololo Inter-American Observatory, NSF’s NOIRLab; the Bok telescope, Steward Observatory, University of Arizona; and the Mayall telescope, Kitt Peak National Observatory, NOIRLab. Pipeline processing and analyses of the data were supported by NOIRLab and the Lawrence Berkeley National Laboratory (LBNL). The Legacy Surveys project is honored to be permitted to conduct astronomical research on I’oligam Du’ag (Kitt Peak), a mountain with particular significance to the Tohono O’odham Nation.

NOIRLab is operated by the Association of Universities for Research in Astronomy (AURA) under a cooperative agreement with the National Science Foundation. LBNL is managed by the Regents of the University of California under contract to the U.S. Department of Energy.

This project used data obtained with the Dark Energy Camera (DECam), which was constructed by the Dark Energy Survey (DES) collaboration. Funding for the DES Projects has been provided by the U.S. Department of Energy, the U.S. National Science Foundation, the Ministry of Science and Education of Spain, the Science and Technology Facilities Council of the United Kingdom, the Higher Education Funding Council for England, the National Center for Supercomputing Applications at the University of Illinois at Urbana-Champaign, the Kavli Institute of Cosmological Physics at the University of Chicago, Center for Cosmology and Astro-Particle Physics at the Ohio State University, the Mitchell Institute for Fundamental Physics and Astronomy at Texas A\&M University, Financiadora de Estudos e Projetos, Fundacao Carlos Chagas Filho de Amparo, Financiadora de Estudos e Projetos, Fundacao Carlos Chagas Filho de Amparo a Pesquisa do Estado do Rio de Janeiro, Conselho Nacional de Desenvolvimento Cientifico e Tecnologico and the Ministerio da Ciencia, Tecnologia e Inovacao, the Deutsche Forschungsgemeinschaft and the Collaborating Institutions in the Dark Energy Survey. The Collaborating Institutions are Argonne National Laboratory, the University of California at Santa Cruz, the University of Cambridge, Centro de Investigaciones Energeticas, Medioambientales y Tecnologicas-Madrid, the University of Chicago, University College London, the DES-Brazil Consortium, the University of Edinburgh, the Eidgenossische Technische Hochschule (ETH) Zurich, Fermi National Accelerator Laboratory, the University of Illinois at Urbana-Champaign, the Institut de Ciencies de l’Espai (IEEC/CSIC), the Institut de Fisica d’Altes Energies, Lawrence Berkeley National Laboratory, the Ludwig Maximilians Universitat Munchen and the associated Excellence Cluster Universe, the University of Michigan, NSF’s NOIRLab, the University of Nottingham, the Ohio State University, the University of Pennsylvania, the University of Portsmouth, SLAC National Accelerator Laboratory, Stanford University, the University of Sussex, and Texas A\&M University.

BASS is a key project of the Telescope Access Program (TAP), which has been funded by the National Astronomical Observatories of China, the Chinese Academy of Sciences (the Strategic Priority Research Program “The Emergence of Cosmological Structures” Grant \# XDB09000000), and the Special Fund for Astronomy from the Ministry of Finance. The BASS is also supported by the External Cooperation Program of Chinese Academy of Sciences (Grant \# 114A11KYSB20160057), and Chinese National Natural Science Foundation (Grant \# 12120101003, \# 11433005).

The Legacy Survey team makes use of data products from the Near-Earth Object Wide-field Infrared Survey Explorer (NEOWISE), which is a project of the Jet Propulsion Laboratory/California Institute of Technology. NEOWISE is funded by the National Aeronautics and Space Administration.

The Legacy Surveys imaging of the DESI footprint is supported by the Director, Office of Science, Office of High Energy Physics of the U.S. Department of Energy under Contract No. DE-AC02-05CH11231, by the National Energy Research Scientific Computing Center, a DOE Office of Science User Facility under the same contract; and by the U.S. National Science Foundation, Division of Astronomical Sciences under Contract No. AST-0950945 to NOAO.

This research is based on observations made with the Galaxy Evolution Explorer (GALEX), obtained from the MAST data archive at the Space Telescope Science Institute, which is operated by the Association of Universities for Research in Astronomy, Inc., under NASA contract NAS5–26555. The GALEX FUV data used in this paper can be found in MAST \dataset[10.17909/048p-bf61]{http://dx.doi.org/10.17909/048p-bf61}.

We also used the NASA/IPAC Extragalactic Database
(NED; \dataset[10.26132/NED1]{https://doi.org/10.26132/NED1}),
which is operated by the Jet Propulsion Laboratory, California Institute of Technology,
under contract with the National Aeronautics and Space Administration.

We acknowledge the use of Claude (Anthropic; model Opus 5.5) in reviewing and debugging the analysis code and improving the language and organization of the text. 

\end{acknowledgments}

\facilities{Keck:II (KCWI/KCRM), Blanco (DECam), GALEX, NED}

\software{Astropy \citep{Astropy_Collaboration:2013,Astropy_Collaboration:2018,Astropy_Collaboration:2022}, NumPy \citep{Harris:2020}, SciPy \citep{Virtanen:2020}, Matplotlib \citep{Hunter:2007}, KCWI DRP \url{https://github.com/Keck-DataReductionPipelines/KCWI_DRP}, CWItools \citep{OSullivan:2020}, VorBin \citep{Cappellari:2012}, pysersic \citep{Pasha:2023}, CRCombine \url{https://github.com/prappleizer/CRCombine}
          }

\bibliography{references}{}
\bibliographystyle{aasjournal}



\end{document}